\documentclass{article}
\usepackage{ijcai24}

\usepackage{times}
\usepackage{soul}
\usepackage{url}
\usepackage[hidelinks]{hyperref}
\usepackage[utf8]{inputenc}
\usepackage[small]{caption}
\usepackage{graphicx}
\usepackage{amsmath}
\usepackage{amsthm}
\usepackage{booktabs}
\usepackage{algorithm}
\usepackage{algorithmic}
\usepackage[switch]{lineno}
\usepackage{tabularx}
\usepackage{multirow}
\usepackage{color}
\usepackage{subcaption}
\usepackage{enumitem}
\usepackage{amsfonts}
\usepackage{pifont}

\title{
Break the ID-Language Barrier: An Adaption Framework for LLM-based Sequential Recommendation
}

\author{
Xiaohan Yu$^1$
\and
Li Zhang $^{2,3}$\footnote{Corresponding Author} \and
Xin Zhao $^1$\and
Yue Wang $^1$\
\\
\affiliations
$^1$ Huawei, Beijing, China\\
$^2$ Institute of Finance Technology, UCL, United Kingdom \\
$^3$ Civil, Environmental and Geomatic Engineering, UCL, United Kingdom\\
\emails
\{yuxiaohan5, zhaoxin151, wangyue262\}@huawei.com,
ucesl07@ucl.ac.uk
}

\begin{document}

\maketitle

\begin{abstract}


The recent breakthrough of large language models (LLMs) in natural language processing has sparked exploration in recommendation systems, however, their limited domain-specific knowledge remains a critical bottleneck. Specifically, LLMs lack key pieces of information crucial for sequential recommendations, such as user behavior patterns. To address this critical gap, we propose IDLE-Adapter, a novel framework that integrates pre-trained ID embeddings, rich in domain-specific knowledge, into LLMs to improve recommendation accuracy. IDLE-Adapter acts as a  bridge, transforming sparse user-item interaction data into dense, LLM-compatible representations through a Pre-trained ID Sequential Model, Dimensionality Alignment, Layer-wise Embedding Refinement, and Layer-wise Distribution Alignment. Furthermore, IDLE-Adapter demonstrates remarkable flexibility by seamlessly integrating ID embeddings from diverse ID-based sequential models and LLM architectures. Extensive experiments across various datasets demonstrate the superiority of IDLE-Adapter, achieving over 10\% and 20\% improvements in HitRate@5 and NDCG@5 metrics, respectively, compared to state-of-the-art methods.


\end{abstract}

\section{Introduction}

Sequential recommendation has emerged as a new paradigm that explicitly models user behavior sequences for near-future recommendations \cite{wang2019sequential}. This paradigm effectively captures the complex dependencies within and between user sequences to accurately infer their evolving preferences. Existing methods, such as Markov Chains, LSTMs, GRUs, CNNs, and self-attention mechanisms, have achieved notable success \cite{rendle2010factorizing,wu2017recurrent,hidasi2015session,caser,kang2018self}. However, their reliance on item identifiers (IDs) presents a critical bottleneck, particularly in data-sparse environments where generalization becomes increasingly challenging.

Recently, large language models (LLMs) have shown impressive compositional and reasoning abilities in a variety of NLP tasks, demonstrating great potential for more intelligent recommendations. Preliminary findings reveal that LLM-based recommendation approaches exhibit significant advantages in cold-start or long-tail scenarios, surpassing traditional sequential models through superior capacity to generalize beyond observed data \cite{gao2023chat,hou2023cold,sanner2023cold}. However, the integration of LLMs into the realm of recommendation systems remains in its nascent stages and faces fundamental challenges. Early exploration, exemplified by P5 \cite{geng2022recommendation} and its variants \cite{hua2023up5,geng2023vip5}, represents users and items with ID numbers (e.g., \textit{Given the user 15's purchase history: 115, 301, 24, predict the next item to be purchased}). While seemingly straightforward, this approach poses a significant limitation, as ID numbers themselves lack inherent semantic meaning, unlike the word tokens, rendering it vulnerable to generalization and transferability. Recent efforts have attempted to address this limitation by transforming user sequential interactions into textual information such as item titles (e.g., \textit{Given the user purchase history: a skirt, a coat,..., predict the next item to be purchased}) \cite{Zhang2021,cui2022m6,wang2023generative} to exploit the semantic richness of LLMs. However, LLMs face inherent limitations in processing lengthy sequences \cite{beltagy2020longformer}, restricting their ability to fully leverage comprehensive user-item interactions. Additionally, relying solely on textual information compromises the level of granularity and precision achievable with explicit ID-based features.

These observations highlight the crucial need to move beyond simple ID strings and integrate the rich domain knowledge embedded within user-item interactions. A natural approach would be direct integration into LLMs, but the disparity between interaction data and LLM-compatible language inputs presents a significant challenge. 
While sequential models excel at capturing essential domain aspects such as user behavior patterns in ID embeddings, they often lack the broader world knowledge and textual understanding that LLMs possess. This motivates our use of a sequential model-based adaptation phase to enrich the LLM-powered recommendation approach.

In this work, we propose an \textbf{ID}-\textbf{L}angaug\textbf{E} \textbf{Adapter} framework (IDLE-Adapter), that aims to seamlessly integrate the strengths of domain-specific knowledge embedded in ID embeddings with the semantic understanding capabilities of LLMs. Our key contribution lies in effectively aligning the latent spaces of ID embeddings and LLMs. ID-based models typically generate sparse representations based on historical interactions, while LLMs operate on dense, contextualized features.
The IDLE-Adapter acts as a bridge, transforming sparse user-item interaction data into dense, LLM-compatible representations through a four-step process: pre-trained ID sequential model, dimensionality alignment, layer-wise embedding refinement and layer-wise distribution alignment. Recognizing that different layers within an LLM attend to diverse aspects of the input features, we propose a novel layer-wise adaptation mechanism to address the dimensionality and distribution mismatch between ID-based and LLM embeddings. Subsequent layer-wise refinement grants fine-grained control over the user embedding influence at each layer. To further brdige the gap, we leverage Maximum Mean Discrepancy minimization to align the underlying latent distributions of ID-based and LLM embeddings.

Extensive evaluation on three real-world public datasets demonstrates the effectiveness of IDLE-Adapter. Compared to state-of-the-art models, IDLE-Adapter achieves significant improvements in both HitRate@5 and NDCG@5 metrics, exceeding them by over 10\% and 20\%, respectively, indicating the efficacy of fusing recommendation domain knowledge encoded in ID representations with the expressive power of LLMs. Moreover, we delve into the generalizability of IDLE-Adapter, proving its ability to seamlessly collaborate with a wide range of ID-based sequential models and LLM backbones. Overall, our contributions are summarized as follows:

\begin{itemize}

    \item We propose a novel framework, IDLE-Adapter, that unlocks the potential of LLMs for recommendation tasks by seamlessly integrating domain-specific knowledge from traditional sequential models. 
    
    \item IDLE-Adapter enables effective collaboration between diverse ID-based models and LLM architectures, showcasing exceptional flexibility and generalizability in integration capabilities. 
    
    \item Extensive experiments on three public datasets demonstrate the superiority of IDLE-Adapter. In addition, ablation experiments and a thorough analysis of the IDLE-Adapter are given.
\end{itemize}

\begin{figure*}[t!]
    \centering
    \includegraphics[width=16cm]{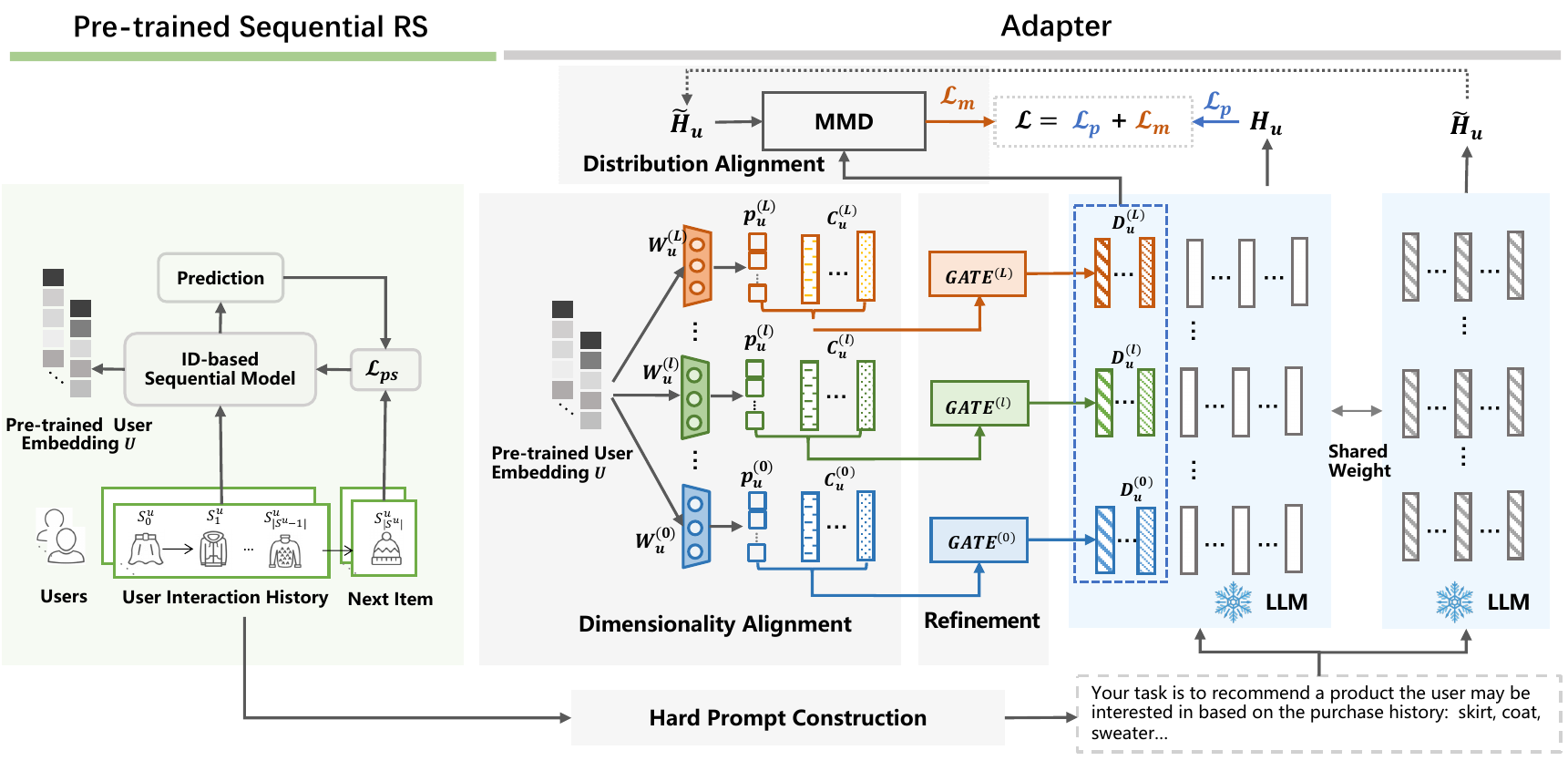}
    \caption{The overall framework of IDLE-Adapter.}
    \label{fig:methodology}
\end{figure*}

\section{Related Work}
\subsection{Sequential Recommendation}

The landscape of sequential recommendation has been enriched by a vibrant tapestry of methodologies. Earlier studies mainly adopt Markov Chain models to learn local transition patterns between items \cite{rendle2010factorizing}. Recurrent architectures, such as GRU4Rec \cite{hidasi2015session}, introduced Gated Recurrent Units to effectively capture sequential patterns within user histories. Convolutional Neural Networks (CNNs) emerged with models like Caser \cite{caser}, utilizing convolutional filters to extract hidden information from user interaction sequences. Recent years have witnessed significant advancements in self-attention mechanisms that have been widely applied in sequential recommendations, as exemplified by SASRec \cite{kang2018self}, which has yielded impressive results add citations. Moreover, self-supervised learning has gained traction, with models like S3Rec \cite{zhou2020s3} and CL4SRec \cite{xie2022contrastive} capitalizing on auxiliary signals to further enhance recommendation accuracy. However, traditional sequential models heavily rely on IDs, which hinders their generalization ability, especially in data-sparse settings.



\subsection{LLMs for Recommendation}
The recent blossoming of large language models has sparked a surge of interest in their application to recommender systems. Capitalizing on their emergent capabilities, LLMs offer a promising avenue for content-based recommendation.
Pioneering works like P5 \cite{geng2022recommendation} directly generate next item IDs, effectively transforming diverse recommendation tasks into language generation tasks. Follow-up works\cite{hua2023up5,geng2023vip5,hua2023index} have delved deeper, exploring multimodal integration and optimized indexing strategies to augment P5's core approach.
Another vein of research seeks to tailor LLMs specifically for recommendation domains. InstructRec \cite{zhang2023instruction} frames recommendations as a series of 39 distinct instructions and employing GPT-3.5 for supervised finetuning. GPT4Rec \cite{li2023gpt4rec} harnesses the power of natural language descriptions, constructing comprehensive prompts that encapsulate user preferences and insights. Despite these significant advances, LLM-based RS are struggling to capture domain-specific knowledge embedded within user and item IDs. Consequently, future research endeavors are needed to develop expressive ID embeddings tailored for LLM-based recommendation systems.

\section{Task Formulation}
Consider a set of $N$ users $\mathcal{U}=\{u_1, u_2, \ldots u_N\}$ interacting with a set of $M$ items $\mathcal{I}=\{i_1, i_2, \ldots i_M\}$. We denote the collection of all user interaction sequences as $\mathcal{S}=\{\mathcal{S}^1, \mathcal{S}^2, \ldots \mathcal{S}^{\left\vert \mathcal{U}\right\vert}\}$, where each $\mathcal{S}^{u}$ represents the chronologically ordered sequence of items interacted with by user $u \in \mathcal{U}$. 
Formally, we define $\mathcal{S}^{u} = (\mathcal{S}^{u}_{0},\mathcal{S}^{u}_{1},\ldots,\mathcal{S}^{u}_{\left\vert S^{u}\right\vert})$, where $\left\vert S^{u} \right\vert$ denotes the sequence length and $\mathcal{S}^{u}_{t} \in \mathcal{I}$ is the item interacted by user $u$ at time $t$. The primary objective of sequential recommendation is to predict the next item $\mathcal{S}^{u}_{\left\vert S^{u}\right\vert+1}$ that user $u$ will interact with.

In this paper, we propose a framework utilizing large language models for sequential recommendation that leverages both textual information and user-item interaction sequences ($\mathcal{S}$). 
Let $T_{ut}$ denote the natural language input associated with user $u$ at time $t$. 
Due to the inherent heterogeneity between text and sequential interaction data (different formats, sparsity, etc.), directly integrating the interaction sequences into the LLM presents a significant challenge.
To address this challenge, we propose a function $\mathcal{F}$ that encapsulates sparse user interaction history into the native space of the LLM. The goal is to approach the ideal $\mathcal{F}$:
\begin{align}
    \mathcal{E}{\phi} &:= E_{(\mathcal{S}^u, t)\sim \mathcal{S}}[l(\mathcal{S}_{t+1}^u, \mathcal{F}(\mathcal{S}_{1:t}^u, T_{ut}))]
\end{align}
where $l(\cdot, \cdot)$ denotes the prediction error function between the prediction and the ground truth.

\section{Our Proposed Framework}
The IDLE-Adapter framework, as depicted in Figure \ref{fig:methodology}, is distinguished by two foundational components: hard prompt design and adapter. Hard prompt design component leverages meticulously constructed prompt templates.
The adapter acts as a bridge, transforming sparse user-item interaction data into dense, LLM-compatible representations through a four-step process: Pre-trained ID Sequential Model, Dimensionality Alignment, Layer-wise Embedding Refinement and Layer-wise Distribution Alignment.

\subsection{Hard Prompt Construction}
The hard prompt $T_{ut}$ aims to convert the user interaction sequences $\mathcal{S}^{u}$ to the natural language and it comprises two key components: explicit directives that define the recommendation objective and informative details about the users and items \cite{li2023gpt4rec,yu2024ra}. 
To establish the high-level goal and context, we incorporate instructions summarizing the task. Furthermore, we enrich the prompt with pertinent details about users (e.g., past purchasing behavior) and items (e.g., titles) to equip LLMs with necessary information for accurate recommendations. 
Figure \ref{fig:enter-label} exemplifies a hard prompt incorporating these elements, \textit{"Your task is to recommend three products the user may be interested in based on their purchase history: skirt, coat, sweater..."}.

\subsection{Adpater}


\subsubsection{Pre-trained ID-based Sequential Model}
Given the set of user interaction sequences $\mathcal{S}$, we establish an item embedding matrix $\mathbf{I} \in \mathbb{R}^{M \times d}$ and a user embedding matrix $\mathbf{U} \in \mathbb{R}^{N \times d}$, where $d$ is the embedding dimension. We apply an ID-based sequential recommendation model $\mathcal{F}':= \mathcal{S} \mapsto \mathbb{R}^{N \times d}$, parameterized by $\Theta_{ID}$, which maps each user with interaction sequence $\mathcal{S}^u= (\mathcal{S}^{u}_{0},\mathcal{S}^{u}_{1},\ldots,\mathcal{S}^{u}_{\left\vert S^{u}-1 \right\vert})$ to a dense embedding vector $\mathbf{u} \in \mathbb{R}^{1 \times d}$ as follows:
\begin{equation}
    \mathbf{u} = \mathcal{F}'(\mathcal{S}^{u}\vert \Theta_{ID}),
\end{equation}
Then, we predict the interaction likelihood (e.g. click or not) between $u$ and item $i$:
\begin{align}
    P_{\Theta_{ID}}(i|u) = \frac{{\rm exp}(\mathbf{u}^T \mathbf{i})}{\sum_{i'\in \mathcal{I}} {\rm exp}(\mathbf{u}^T \mathbf{i'})},
\end{align}
where $\mathbf{i} \in \mathbb{R}^{1 \times d}$ is the item embedding vector. Following the paradigm of classical sequential recommendation models \cite{hidasi2015session,cen2020controllable}, we employ cross entropy loss to optimize our pre-trained model, formulated as:
\begin{align}
    &\mathcal{L}_{ps} = -\sum_{u\in \mathcal{U}} \sum_{i\in \mathcal{I}} {\rm log} P_{\Theta_{ID}}(i|u),
\end{align}

\subsubsection{Dimensionality Alignment}
The integration of pre-trained ID base model with large language models poses a substantial challenge in the form of dimensionality mismatch.
Typically, pre-trained user embeddings typically reside in a lower-dimensional space compared to the internal representations manipulated by LLMs. To bridge this gap and ensure compatibility, we propose to dynamically align the dimension of the pre-trained user embedding $\mathbf{u}$ with each corresponding layer of the LLM, denoted as $\mathbb{R}^d \mapsto \mathbb{R}^{d'}$.

\subsubsection{Layer-wise Embedding Refinement}
With the dimensionally aligned user representations, we propose to integrate them into each layer of the LLMs as \textit{virtual tokens} \cite{li2021prefix}. However, existing research \cite{jawahar-etal-2019-bert,tenney2019bert} demonstrates that different layers within a language model attend to distinct aspects of the input features. 
This heterogeneity necessitates a mechanism for adapting user embeddings at each layer, granting the model fine-grained control over the influence of the incorporated user representations across layers. 
Building upon the success of recent reparameterization techniques \cite{hu2021lora,edalati2022krona,aghajanyan2020intrinsic}, we introduce a layer-specific user projection:
\begin{equation}
\mathbf{p}_{u}^{(l)} = \mathbf{W}_{u}^{(l)} \mathbf{u} + \mathbf{b}_u^{(l)},
\end{equation}
where $l\in[0, L]$ denotes the layer index and $\mathbf{W}_{u}^{(l)} \in \mathbb{R}^{d'\times d},\mathbf{b}_{u}^{(l)}\in \mathbb{R}^{d'}$ are trainable parameters. 
To further refine the user embeddings within each layer, we introduce a layer-specific refinement mechanism based on the layer-specific prefix $\mathbf{C}_u^{(l)} \in \mathbb{R}^{L' \times d'}$, where  $L'$ denotes the length of the integrated \textit{virtual tokens}.
These prefixes delve into the token level, offering fine-grained control over the alignment process. To determine the necessary adjustment for each layer's prompt, we employ a token-level combination gate mechanism, represented as $\mathbf{g}_u^{(l)} \in \mathbb{R}^{L'}$:
\begin{align}
    \mathbf{g}_u^{(l)} = \sigma ({\rm Concat}(\mathbf{C}_u^{(l)} , \mathbf{P}_u^{(l)} )\mathbf{W}_{g}),
\end{align}
where $\sigma(\cdot)$ represents the sigmoid function, $\mathbf{W}_{g} \in \mathbb{R}^{2d'}$ is a learnable matrix and $\mathbf{P}_u^{(l)} \in \mathbb{R}^{L' \times d'} $ is the repetition of $\mathbf{p}_u^{(l)}$ by $L'$ times. $\mathbf{g}_u^{(l)}$ acts as a blend factor, dynamically controlling the influence of the layer-specific prefix 
$\mathbf{C}_u^{(l)}$ and the user representation $\mathbf{P}_u^{(l)}$. 
We formulate the refined user representation:
\begin{align}
\mathbf{D}_u^{(l)} = \mathbf{g}_u^{(l)} \odot \mathbf{C}_u^{(l)} + (1- \mathbf{g}_u^{(l)}) \odot \mathbf{P}_u^{(l)},
\label{soft_prompt}
\end{align}
where $\odot$ is the element-wise multiplication. The refined user representation $\mathbf{D}_u^{(l)}$ is then integrated into the self-attention mechanism within each layer as follows:
\begin{align}
{\rm SA(\mathbf{Q}, \mathbf{K}, \mathbf{V})} &= {\rm Softmax}(\frac{\mathbf{Q}^T\mathbf{K}}{\sqrt{d'}})\mathbf{V}.
\end{align}
Crucially, the key and value matrices are modified to incorporate both the hard prompt and the refined user representations:
\begin{align}
\mathbf{Q} &= \mathbf{W}_q^T \mathbf{H}_u^{(l)}, \\
\mathbf{K} &= \mathbf{W}_k^T {\rm Concat}(\mathbf{D}_u^{(l)}, \mathbf{H}_u^{(l)}), \\
\mathbf{V} &= \mathbf{W}_v^T {\rm Concat}(\mathbf{D}_u^{(l)}, \mathbf{H}_u^{(l)}),
\end{align}
where $\mathbf{H}_u^{(l)}$ denotes the hidden states at layer $l$ of large language model, $\mathbf{W}_k \in \mathbb{R}^{d'\times 2d'}, \mathbf{W}_v \in \mathbb{R}^{d'\times 2d'}, \mathbf{W}_q \in \mathbb{R}^{d'\times d'} $ represent the transformations applied to the keys, values and queries, respectively.

\subsubsection{Layer-wise Distribution Alignment}
To further bridge the gap between the representations learned from ID-based sequential recommendation and pre-trained language models, we posit that minimizing the divergence between their underlying latent distributions is crucial. Drawing inspiration from domain adaptation techniques, we propose the layer-wise distribution alignment module, which directly operates on the intermediate layers of both the sequential recommendation and LLM encoders.

To achieve this alignment, we leverage the Maximum Mean Discrepancy (MMD) \cite{gretton2006kernel}, a principled measure of distribution discrepancy operating in a reproducing kernel Hilbert space. In our specific application, we minimize the MMD between the adapted user representation $\mathbf{D}_u$ (obtained from Eq. \ref{soft_prompt}) and the corresponding LLM representations, denoted as $\mathbf{\Tilde{H}}_u$.
The LLM representations are obtained from the same underlying language model but without incorporating the adapted user representation as input at any layer.
The MMD loss for layer-wise alignment is formulated as:
\begin{equation}
\begin{split}
    \mathcal{L}_{m} & = M_k(\mathbf{D}_u, \mathbf{\Tilde{H}}_u) = \frac{1}{L^2} \sum_{a=1}^L \sum_{b=1}^L \mathcal{K}(\mathbf{d}_a, \mathbf{d}_b) \\
    & + \frac{1}{L^2} \sum_{a=1}^L \sum_{b=1}^L \mathcal K(\mathbf{\Tilde{h}}_a, \mathbf{\Tilde{h}}_b) - \frac{2}{L^2} \sum_{a=1}^L \sum_{b=1}^L \mathcal{K} (\mathbf{d}_a, \mathbf{\Tilde{h}}_b),
    \label{align}
\end{split}
\end{equation}
where $a, b$ denotes the row index and $\mathcal{K}(\cdot, \cdot)$ is a chosen kernel function, Gaussian kernel, i.e., $\mathcal{K}(x, y)=e^{-\frac{||x-y||^2}{2\rho}}$, and $\rho$ is a predefined parameter.

\subsection{Joint Optimization}
We jointly train the model via two objectives, next item prediction and the distribution alignment. Given the user interaction history sequence $\mathcal{S}^{u} = (\mathcal{S}^{u}_{0},\mathcal{S}^{u}_{1},\ldots,\mathcal{S}^{u}_{\left\vert S^{u}\right\vert})$, we aim to predict the next item the user will interact with. Following \cite{hidasi2015session}, we split the sequence into a set of subsequences with shifted target labels, ${(\mathcal{S}^{u}_{0}, \mathcal{S}^{u}_{1}), (\mathcal{S}^{u}_{0:1}, \mathcal{S}^{u}_{2}), \ldots (\mathcal{S}^{u}_{0:\left\vert S^{u}-1\right\vert}, \mathcal{S}^{u}_{\left\vert S^{u}\right\vert})}$. Then, we formulate our main learning objective based on cross-entropy loss:

\begin{align}
    \mathcal{L}_{p} = -\sum_{\mathcal{S}^u \in \mathcal{S}}\sum_{k=0}^{|\mathcal{S}^u|-1} \log(\hat{\mathbf{y}}^{\mathcal{S}^u_{0:k}}(\mathcal{S}^{u}_{k+1})),
    \label{predict}
\end{align}
where $\hat{\mathbf{y}}^{\mathcal{S}^u_{0:k}}$ represents the model prediction probability distribution over item candidates, conditioned on the user's interaction history $S^u_{0:k}$, defined as:

\begin{align}
   \hat{\mathbf{y}}^{\mathcal{S}} = {\rm Softmax}(\mathbf{W}_y^T \mathbf{H}_u^{(L)}),
\end{align}
where $\hat{\mathbf{y}}^{\mathcal{S}} \in \mathbb{R}^{M\times 1} $,  $\mathbf{W}_y \in \mathbb{R}^{d' \times M}$ is the learnable matrix for the prediction layer, $\mathbf{H}^{(L)}$ is the hidden state at the last layer of LLM. 
The final training objective is as follows:
\begin{equation}
    \mathcal{L} = \mathcal{L}_{p} + \lambda\mathcal{L}_{m},
\end{equation}
where $\lambda$ is a hyperparameter to control the contribution alignment loss towards the overall objective.

To prioritize cost-effectiveness and avoid impacting the LLM's performance, we employ a parameter freezing strategy. We solely update the parameters of the adapter module, while keeping the LLM and the ID-based recommendation model's parameters fixed during fine-tuning. Formally, the gradient update is restricted to:
\begin{align}
    \Delta \Theta_{A} \xleftarrow{} \nabla_{\Theta_{A}} \mathcal{L}(\Theta_{LLM}, \Theta_{ID}, \Theta_{A}; \mathcal{S}),
\end{align}
where $\Theta_{A}$ contains parameters from the adapter module, $\Theta_{ID}$ and $\Theta_{LLM}$ represent the frozen parameters of the ID-based model and the LLM, respectively.

\section{Experiment}

\subsection{Experimental Setup}

\begin{table}[]
\centering
\begin{tabular}{ccccc}
\toprule[1.5pt]
Dataset & \#Users  & \#Items   & \#Interactions & \#Density\\ \midrule
ML-1M    & 6,040 & 3,706 & 1,000,209 &  4.4684\%     \\
Musical   & 26,397 & 53,905 & 274,878 &  0.0193\%     \\
Clothing   & 50,000 & 270,563 & 562,371 &  0.0041\%     \\
\bottomrule[1.5pt]
\end{tabular}
\caption{Statistics of the benchmark datasets.}
\label{data}
\end{table}

\begingroup
\setlength{\tabcolsep}{2.8pt} 
\renewcommand{\arraystretch}{1} 
\begin{table*}[!t]
\centering
\begin{tabular}{l | c c c c | c c c c | c c c c}
\toprule[1.5pt]
\multirow{2}{*}{Models} & \multicolumn{4}{c|}{Clothing} & \multicolumn{4}{c|}{Musical} & \multicolumn{4}{c}{ML-1M} \\ \cmidrule{2-13}
 & HR@5  & HR@10  & N@5 & N@10 & HR@5 & HR@10 & N@5 & N@10 & HR@5 & HR@10 & N@5 & N@10  \\ \midrule
FPMC & 0.0163 & 0.0232 & 0.0089 & 0.0112 & 0.0055 & 0.0168 & 0.0023 & 0.0060 & 0.0084 & 0.0197 & 0.0033 & 0.0080 \\
BPRMF & 0.0093 & 0.0365 & 0.0036 & 0.0124 & 0.0105 & 0.0336 & 0.0041 & 0.0117 & 0.0061 & 0.0182 & 0.0032 & 0.0071 \\
Caser & 0.0417 & 0.0456 & \underline{0.0365} & \underline{0.0377} & 0.0480 & 0.0567 & 0.0394 & 0.0422 & 0.0861 & 0.1546 & 0.0514 & 0.0733 \\
GRU4Rec & 0.0455 & 0.0517 & 0.0296 & 0.0317 & 0.0514 & 0.0656 & \underline{0.0397} & \underline{0.0442} & 0.0465 & 0.0882 & 0.0277 & 0.0411 \\
SASRec & 0.0301 & 0.0436 & 0.0188 & 0.0233 & 0.0417 & 0.0530 & 0.0322 & 0.0355 & 0.0874 & 0.1596 & 0.0503 & 0.0734  \\
ComiRec & \underline{0.0564} & \underline{0.0640} & 0.0301 & 0.0326 & \underline{0.0606} & \underline{0.0752} & 0.0375 & 0.0422 & 0.0727 & 0.1546 & 0.0377 & 0.0638 \\ 
Bert4Rec & 0.0248 & 0.0371 & 0.0155 & 0.0195 & 0.0338 & 0.0442 & 0.0197 & 0.0224 & \underline{0.1016} & \underline{0.1702} & \underline{0.0638} & \underline{0.0868}  \\
S$^3$-Rec & 0.0133 & 0.0174 & 0.0075 & 0.0088 & 0.0402 & 0.0516 & 0.0311 & 0.0347 & 0.0253 &  0.0400 & 0.0176 & 0.0223 \\
\midrule

P5 & 0.0290 & 0.0344 & 0.0252 & 0.0207 & 0.0214 & 0.0349 & 0.0132 & 0.0175 & 0.0267 & 0.0331 & 0.0240 & 0.0260 \\
TwinBert & 0.0310 & 0.0494 & 0.0188 & 0.0247 & 0.0446 & 0.0512 & 0.0390 & 0.0411 & 0.0660 & 0.1242 & 0.0384 & 0.0570 \\
GPT4Rec & 0.0305 & 0.0350 & 0.0264 & 0.0278 & 0.0125 & 0.0220 & 0.0071 & 0.0102 & 0.0050 & 0.0080 & 0.0022 & 0.0032 \\ 
\midrule
IDLE-Adapter & \textbf{0.0656} & \textbf{0.0687} & \textbf{0.0542} & \textbf{0.0552} & \textbf{0.0679} & \textbf{0.0838} & \textbf{0.0580} & \textbf{0.0631} & \textbf{0.1321} & \textbf{0.2056} & \textbf{0.0880} & \textbf{0.1116} \\
Improv.(\%) &  14.02\% & 7.34\% & 48.49\% & 46.42\% & 12.05\% & 11.44\% & 46.09\% & 42.76\% & 20.79\% & 37.93\% & 20.79\% &  28.57\% \\ 
\bottomrule[1.5pt]

\end{tabular}
\caption{Overall performance of different recommendation approaches on three benchmark datasets. The best and second-best results are highlighted with \textbf{boldface} and \underline{underlined}. 
}
\label{table:main}
\end{table*}

\subsubsection{Datasets}
We conduct experiments on two benchmark datasets - the Amazon review dataset \cite{ni-etal-2019-justifying} and MovieLens dataset \cite{10.1145/2827872}. For Amazon dataset, we use the subsets Clothing Shoes and Jewelry and Musical Instruments. For MovieLens, we use the ML-1M version for evaluation.
During data preprocessing, we discard invalid samples with empty titles or titles exceeding 200 characters in length. To guarantee each user has sufficient interactions, we filter out users with fewer than 5 interactions. After preprocessing, we randomly sample 50,000 users from the Clothing dataset to create our final evaluation set. 
Table \ref{data} summarizes key statistics of the final preprocessed datasets.
Both datasets are split into train, validation and test sets with 80\%, 10\%, 10\% ratio respectively.

\subsubsection{Evaluation Metrics}
Following~\cite{hidasi2015session,cen2020controllable}, we evaluate sequential recommendation performance using a leave-one-out strategy. For each user's interaction sequence,  we hold out the final two items for validation and testing respectively. The remaining items in the sequence are used to train the model.
To quantify performance, we report Hit Ratio@K and NDCG@K \cite{NDCG} (respectively denoted by HR@K and N@K) where K$\in \{5,10\}$. 

\subsubsection{Baseline}
We compare our proposed method against two categories of baseline models: ID-based methods and LLM-based methods. 
(i) ID-based Methods.
BPRMF \cite{rendle2012bpr} is a matrix factorization method 
with Bayesian personalized ranking loss.
FPMC \cite{rendle2010factorizing} utilizes Markov Chains to learn item transitions.
GRU4Rec \cite{hidasi2015session} employs recurrent neural networks to capture user behavior sequences.
SasRec \cite{kang2018self}, ComiRec \cite{cen2020controllable} and Bert4Rec \cite{sun2019bert4rec} extract information from user sequences with transformer encoders.
S$^3$-Rec \cite{zhou2020s3} applies a self-supervised learning method.
(ii) LLM-based Methods.
P5 \cite{geng2022recommendation} utilizes a fine-tuned FlanT5 \cite{chung2022scaling} to generate ID strings for recommended items.
TwinBert \cite{lu2020twinbert} decouples the representations of query and document with BERT and a crossing layer. 
GPT4Rec \cite{li2023gpt4rec} performs the next item title generation task.

\subsubsection{Implementation Details}
We implement all evaluation methods using Pytorch.
We utilize ComiRec to obtain pre-trained user embeddings. In ComiRec, we set the number of interests to 1 and the embedding size to 64. Maximum sequence lengths are set to 20 for Amazon datasets and 50 for ML-1M.
All parameters in IDLE-Adapter are initialized using a truncated normal distribution in the range $[-0.02, 0.02]$. 
The number of MMD kernels are set to 1.
We employ AdamW optimizer with an initial learning rate of $5\times 10^{-4}$ and a linear decay schedule. The hyperparameter $\lambda$ is searched from $\{0.1, 0.2, 0.5, 1.0\}$.
The backbone language model is chosen from SentenceBert (as shown in Table \ref{table:main}), BGE and OPT.

\subsection{Overall Performance}
The performance comparison results between different methods are summarized in Table \ref{table:main}. We have the following observations:
1) Among ID-based models, ComiRec and Bert4Rec achieve top performance on Amazon and ML-1M datasets, respectively. This highlights the effectiveness of transformer encoders in capturing the dynamics of user interaction history. Furthermore, ID-based models consistently outperform LLM-based models across key metrics. This suggests that the collaborative filtering techniques employed by ID-based sequential models effectively exploit user sequential patterns, extracting crucial information for characterizing user preferences and intentions. Notably, the relatively weak performance of S$^3$-Rec may be caused by different dataset settings, namely, a higher portion of long-tail users and items, as also observed in \cite{recformer}.  However, the strong performance of certain LLM-based models, such as TwinBert exceeding SASRec on some metrics, underscores the potential of language models in recommendation systems. 
2) Our proposed framework, IDLE-Adapter, consistently outperforms both ID-based and LLM-based models, across all evaluation metrics. This compelling evidence validates the effectiveness of our proposed framework. Notably, IDLE-Adapter achieves maximum relative improvements exceeding 20.79\% and 48.49\% in terms of HR@5 and NDCG@5, respectively, compared to the strongest baseline. The significant improvement can be primarily attributed to the effective fusion of recommendation domain knowledge embedded in ID representations with the expressive power of LLMs. Our adapter mechanism effectively aligns the user representation with each layer of the LLM, providing key information for accurate recommendations.

\begin{figure}[!t]
	\centering
	\begin{subfigure}[b]{0.45\textwidth}
		\includegraphics[width=\textwidth]{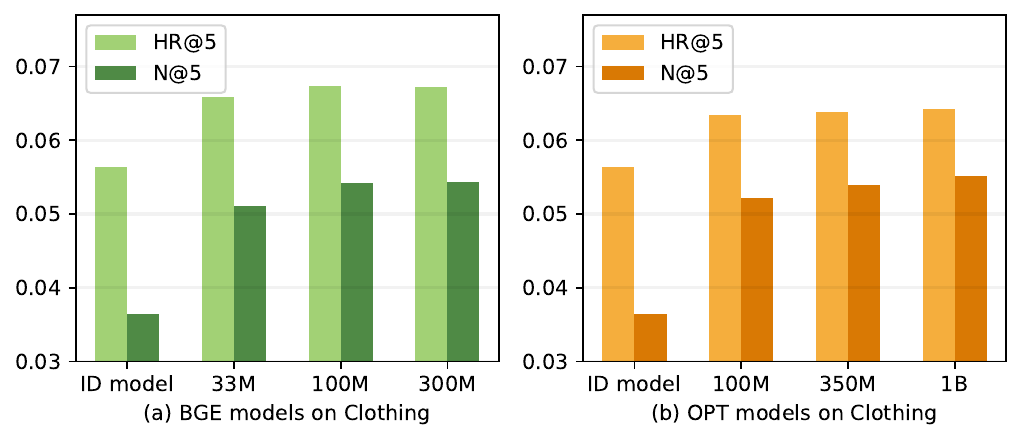}
		\label{fig: loss and step}
	\end{subfigure}
	\begin{subfigure}[b]{0.45\textwidth}
		\includegraphics[width=\textwidth]{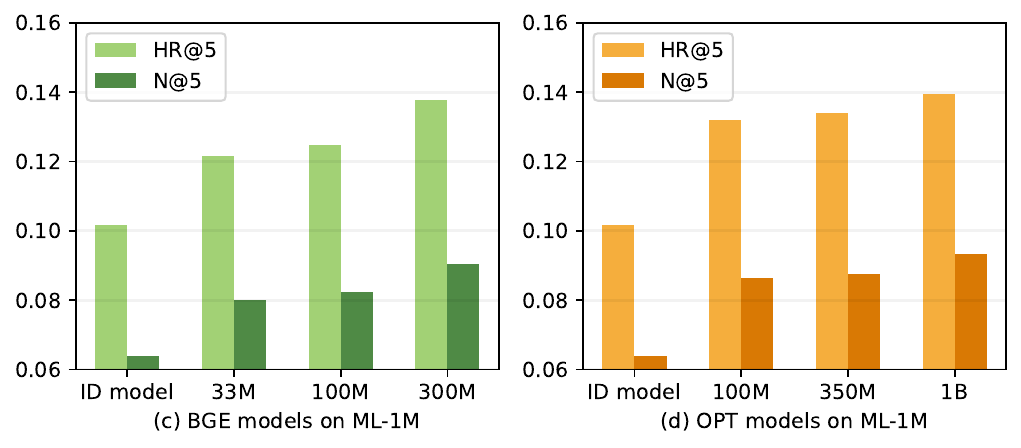}
		\label{fig: hit and step}
	\end{subfigure}
        \caption{Generalizability evaluation of IDLE-Adapter across two series of language models, BGE and OPT. The performance of the second-best ID base model in Table \ref{table:main} is shown on the figure for reference.}
        \label{fig:enter-label}
\label{fig:generalizability}
\end{figure}

\setlength{\tabcolsep}{3pt} 
\begin{table}[!t]
\centering
\begin{tabular}{l | c | c c | c c}
\toprule[1.5pt]
\multirow{2}{*}{ID Model} & \multirow{2}{*}{With LLM?} & \multicolumn{2}{c|}{Musical} & \multicolumn{2}{c}{ML-1M} \\  \cmidrule{3-6}
&  & HR@5  & N@5 & HR@5 & N@5 \\  \cmidrule{1-6}
\multirow{3}{*}{SASRec} & NO & 0.0417 & 0.0322 & 0.0874 & 0.0503 \\ 
 & YES & {0.0563} & 0.0464 & {0.1081} & {0.0657} \\ 
& Improv.(\%)  & 35.0\% & 44.1\% & 23.7\% & 30.6\% \\  \midrule

\multirow{3}{*}{GRU4Rec} & NO & 0.0514 & 0.0397 & 0.0465 & 0.0277 \\ 
& YES  & 0.0631 & {0.0522} & 0.1011 & 0.0670 \\
& Improv.(\%) & 22.8\% & 31.5\% & 117.4\% &  141.9\% \\ 

\bottomrule[1.5pt]
\end{tabular}
\caption{Generalizability evaluation of IDLE-Adapter across diverse pre-trained ID base models. "With LLM" refers to the optimal configuration achieved when pairing the ID base model with any of the considered LLMs (SentenceBERT, BGE, OPT).}
\label{table:generalizability}
\end{table}



\subsection{Generalization Study}
To systematically analyze the generalizability of the IDLE-Adapter framework, we conduct experiments on selection of both large language models and pre-trained ID base models.
\paragraph{Choice of large language models}
We conducted a systematic investigation into the influence of LLM size and architecture on model efficacy. We employ the IDLE-Adapter framework across two distinct LLM architectures: BGE (encoder-based) \cite{bge_embedding} and OPT (decoder-based) \cite{zhang2022opt}. We select models with a range of sizes for each family, with BGE ranging from 33M to 300M parameters and OPT encompassing 100M to 1.3B parameters.
Experiments are conducted on Amazon Clothing and ML-1M. The detailed results are presented in Figure \ref{fig:generalizability}, along with the following key observations:
(1) Performance Scales with LLM Size: A clear correlation emerges between the model size and the efficacy of IDLE-Adapter. As the scale of the language model expands, the performance of the IDLE-Adapter framework gracefully ascends above the baseline ID model. This ascent is a testament to the vastness of world knowledge woven into larger language models, coupled with their enhanced ability to seamlessly fuse semantic and collaborative information.
(2) Encoder vs. Decoder Performance: Holding the model size constant, the encoder-based BGE architecture demonstrates superior performance compared to the decoder-based OPT counterpart on the Amazon Clothing dataset. Conversely, OPT outperforms BGE on ML-1M. This intriguing finding suggests that the decoder model might be more proficient in content-centric scenarios such as movie recommendations, while the encoder model shines in tasks involving product-specific purchase patterns such as clothing recommendation.
\paragraph{Choice of pre-trained ID-based models}
To evaluate the generalizability of our proposed framework across ID-based model architectures, we integrate it with two existing base models, SasRec and MLP4Rec. The results are presented in Table \ref{table:generalizability}. The observed significant improvements on all metrics across different base models validate the generalizability of our framework and its effectiveness in enhancing the performance of ID-based recommender systems. Notably, ComiRec emerges as the optimal choice for generating pre-trained user embeddings, yielding the highest performance.


\subsection{Ablation Study}
To assess the individual contributions of each component in IDLE-Adapter, we conduct ablation study on Amazon Clothing and ML-1M datasets, which are the dimensionality alignment, refinement and distribution alignment modules. The results are shown in Table \ref{ablation}. The detailed configurations of the ablated variants are as follows:
(1) w/o LayerWise: To evaluate the necessity of the fine-grained layer-wise adaptation design in IDLE-Adapter, we replace it with a single uniform adapter, which applies the same adaption step to all LLM layers, disregarding the potentially critical semantic shifts across layers. We observe a significant performance drop compared to the full IDLE-Adapter model, which confirms our hypothesis that explicitly capturing the layer-wise semantic nuances is crucial for effective recommendations. The layer-wise design of adapter enables IDLE-Adapter to tailor user representations specifically for each LLM layer.
(2) w/o Refinement: To evaluate the effectiveness of the refinement module, we remove it entirely, relying solely on dimensionality alignment for user representations adaption. This ablation also led to a performance decline, underscoring the value of the refinement mechanism in enhancing user representation quality.
(3) w/o Distribution: We retain the main objective but eliminated the MMD loss. The observed performance degradation suggests that aligning distribution is a crucial step to bridge the gap between user embeddings and LLMs. 

\setlength{\tabcolsep}{6pt} 
\begin{table}[]
\centering
\begin{tabular}{l | c c | c c}
\toprule[1.5pt]
\multirow{2}{*}{Models} & \multicolumn{2}{c|}{Clothing} & \multicolumn{2}{c}{ML-1M} \\ \cmidrule{2-5}
& HR@5  & N@5 & HR@5  & N@5   \\ \midrule
IDLE-Adapter   & \textbf{0.0656} & \textbf{0.0542} & \textbf{0.1321} & \textbf{0.0880}  \\ \midrule
w/o LayerWise & 0.0599 & 0.0485 & 0.1154 & 0.0758  \\ 
w/o Refinement  & 0.0624 & 0.0512 & 0.1210 & \underline{0.0792} \\ 
w/o Distribution & \underline{0.0626} & \underline{0.0526} & \underline{0.1311} & 0.0782  \\ 
\bottomrule[1.5pt]
\end{tabular}
\caption{Ablation study of IDLE-Adapter on Amazon Clothing and ML-1M.}
\label{ablation}
\end{table}

\begin{figure}[!t]
	\centering
	\begin{subfigure}[b]{0.235\textwidth}
		\includegraphics[width=\textwidth]{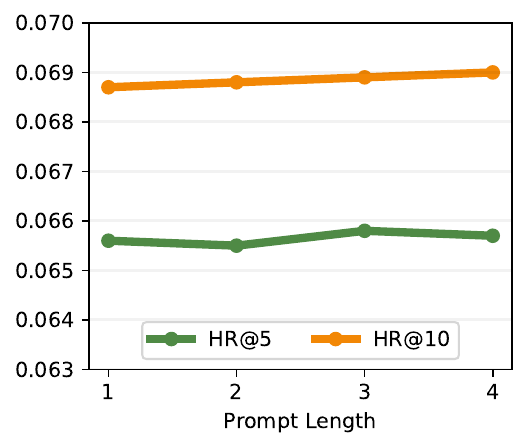}
    \caption{Amazon Clothing}
	\end{subfigure}
	\begin{subfigure}[b]{0.23\textwidth}
		\includegraphics[width=\textwidth]{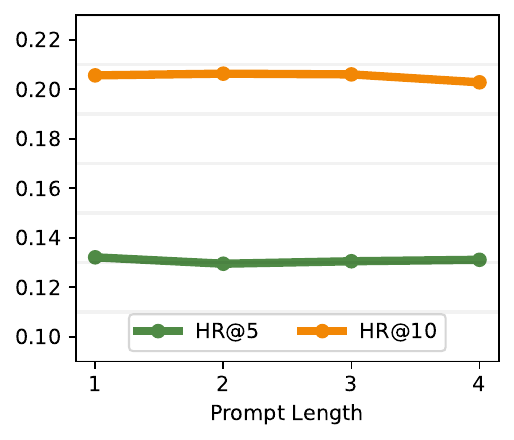}
    \caption{ML-1M}
	\end{subfigure}
\caption{Sensitivity of prompt length.}
\label{fig:hyp}
\end{figure}

\subsection{ Hyperparameter Sensitivity}
We examine the sensitivity of the IDLE-Adapter's performance to the prompt length, a key hyperparameter of the adapter module. We conduct experiments on Amazon Clothing and ML-1M datasets with prompt lengths ranging from 1 to 4. 
As depicted in Figure \ref{fig:hyp}, increasing prompt length initially boosted HitRate@10 on Amazon Clothing, but exhibited fluctuating trends on other settings. This suggests that the effectiveness of our framework is not dependent on prompt length, demonstrating its robustness. The prompt lengths within the 1-2 range appear sufficient for both datasets.

\begin{figure}[t!]
    \centering
    \includegraphics[width=0.3\textwidth]{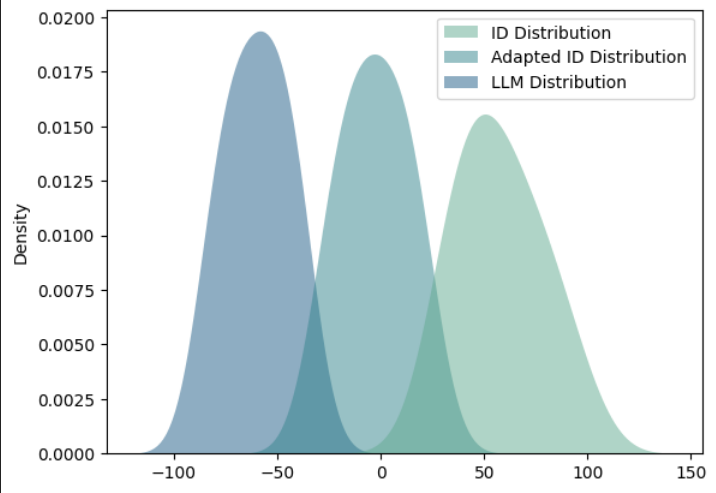}
    \caption{Visualization of the learned representations of the ID base model, the LLM and the adapter.}
    \label{fig:visual}
\end{figure}

\subsection{Explainability of Adapter}
To gain a deeper qualitative understanding of the adapter module, we visualize the learned representations of the ID base model, the LLM backbone model, and the adapter in Figure \ref{fig:visual}. We plot the embedding distribution of these representations with Gaussian kernel density estimation (KDE). The visualization reveals that the adapted ID representations predominantly reside in the region between the ID base model and the LLM, effectively bridging the gap between their feature spaces. Notably, the adapted ID representation distribution exhibits a closer proximity to the LLM's native space compared to the original ID base model. This observation empirically validates the effectiveness of our proposed adapter framework in aligning the representations across ID and language.



\section{Conclusion}
In this paper, we investigate how to integrate the domain-specific knowledge from  ID-based sequential models into LLMs for more powerful and intelligent recommender systems. To bridge the gap caused by different modalities, we propose a novel adaptation framework: IDLE-Adapter that transforms sparse user-item interaction data into dense, LLM-compatible representations with four steps: Pre-trained ID Sequential Model, Dimensionality Alignment, Layer-wise Embedding Refinement, and Layer-wise Distribution Alignment. 
Furthermore, IDLE-Adapter demonstrates remarkable generalization ability and flexibility on diverse ID-based sequential models and LLM architectures. Experimental results demonstrate the effectiveness of the our framework on benchmark datasets.
Additionally, the component-wise effects of our framework are evaluated with ablation study. 

\bibliographystyle{named}
\bibliography{ijcai24}

\begin{thebibliography}{}

\bibitem[\protect\citeauthoryear{Aghajanyan \bgroup \em et al.\egroup }{2020}]{aghajanyan2020intrinsic}
Armen Aghajanyan, Luke Zettlemoyer, and Sonal Gupta.
\newblock Intrinsic dimensionality explains the effectiveness of language model fine-tuning.
\newblock {\em arXiv preprint arXiv:2012.13255}, 2020.

\bibitem[\protect\citeauthoryear{Beltagy \bgroup \em et al.\egroup }{2020}]{beltagy2020longformer}
Iz~Beltagy, Matthew~E Peters, and Arman Cohan.
\newblock Longformer: The long-document transformer.
\newblock {\em arXiv preprint arXiv:2004.05150}, 2020.

\bibitem[\protect\citeauthoryear{Cen \bgroup \em et al.\egroup }{2020}]{cen2020controllable}
Yukuo Cen, Jianwei Zhang, Xu~Zou, Chang Zhou, Hongxia Yang, and Jie Tang.
\newblock Controllable multi-interest framework for recommendation.
\newblock In {\em SIGKDD}, 2020.

\bibitem[\protect\citeauthoryear{Chung \bgroup \em et al.\egroup }{2022}]{chung2022scaling}
Hyung~Won Chung, Le~Hou, Shayne Longpre, Barret Zoph, Yi~Tay, William Fedus, Eric Li, Xuezhi Wang, Mostafa Dehghani, Siddhartha Brahma, et~al.
\newblock Scaling instruction-finetuned language models.
\newblock {\em arXiv preprint arXiv:2210.11416}, 2022.

\bibitem[\protect\citeauthoryear{Cui \bgroup \em et al.\egroup }{2022}]{cui2022m6}
Zeyu Cui, Jianxin Ma, Chang Zhou, Jingren Zhou, and Hongxia Yang.
\newblock M6-rec: Generative pretrained language models are open-ended recommender systems.
\newblock {\em arXiv preprint arXiv:2205.08084}, 2022.

\bibitem[\protect\citeauthoryear{Edalati \bgroup \em et al.\egroup }{2022}]{edalati2022krona}
Ali Edalati, Marzieh Tahaei, Ivan Kobyzev, Vahid~Partovi Nia, James~J Clark, and Mehdi Rezagholizadeh.
\newblock Krona: Parameter efficient tuning with kronecker adapter.
\newblock {\em arXiv preprint arXiv:2212.10650}, 2022.

\bibitem[\protect\citeauthoryear{Gao \bgroup \em et al.\egroup }{2023}]{gao2023chat}
Yunfan Gao, Tao Sheng, Youlin Xiang, Yun Xiong, Haofen Wang, and Jiawei Zhang.
\newblock Chat-rec: Towards interactive and explainable llms-augmented recommender system.
\newblock {\em arXiv preprint arXiv:2303.14524}, 2023.

\bibitem[\protect\citeauthoryear{Geng \bgroup \em et al.\egroup }{2022}]{geng2022recommendation}
Shijie Geng, Shuchang Liu, Zuohui Fu, Yingqiang Ge, and Yongfeng Zhang.
\newblock Recommendation as language processing (rlp): A unified pretrain, personalized prompt \& predict paradigm (p5).
\newblock In {\em Proceedings of the 16th ACM Conference on Recommender Systems}, pages 299--315, 2022.

\bibitem[\protect\citeauthoryear{Geng \bgroup \em et al.\egroup }{2023}]{geng2023vip5}
Shijie Geng, Juntao Tan, Shuchang Liu, Zuohui Fu, and Yongfeng Zhang.
\newblock Vip5: Towards multimodal foundation models for recommendation.
\newblock {\em arXiv preprint arXiv:2305.14302}, 2023.

\bibitem[\protect\citeauthoryear{Gretton \bgroup \em et al.\egroup }{2006}]{gretton2006kernel}
Arthur Gretton, Karsten Borgwardt, Malte Rasch, Bernhard Sch{\"o}lkopf, and Alex Smola.
\newblock A kernel method for the two-sample-problem.
\newblock {\em Advances in neural information processing systems}, 19, 2006.

\bibitem[\protect\citeauthoryear{Harper and Konstan}{2015}]{10.1145/2827872}
F.~Maxwell Harper and Joseph~A. Konstan.
\newblock The movielens datasets: History and context.
\newblock {\em ACM Trans. Interact. Intell. Syst.}, 5(4), dec 2015.

\bibitem[\protect\citeauthoryear{Hidasi \bgroup \em et al.\egroup }{2015}]{hidasi2015session}
Bal{\'a}zs Hidasi, Alexandros Karatzoglou, Linas Baltrunas, and Domonkos Tikk.
\newblock Session-based recommendations with recurrent neural networks.
\newblock {\em arXiv preprint arXiv:1511.06939}, 2015.

\bibitem[\protect\citeauthoryear{Hou \bgroup \em et al.\egroup }{2023}]{hou2023cold}
Yupeng Hou, Junjie Zhang, Zihan Lin, Hongyu Lu, Ruobing Xie, Julian McAuley, and Wayne~Xin Zhao.
\newblock Large language models are zero-shot rankers for recommender systems, 2023.

\bibitem[\protect\citeauthoryear{Hu \bgroup \em et al.\egroup }{2021}]{hu2021lora}
Edward~J Hu, Yelong Shen, Phillip Wallis, Zeyuan Allen-Zhu, Yuanzhi Li, Shean Wang, Lu~Wang, and Weizhu Chen.
\newblock Lora: Low-rank adaptation of large language models.
\newblock {\em arXiv preprint arXiv:2106.09685}, 2021.

\bibitem[\protect\citeauthoryear{Hua \bgroup \em et al.\egroup }{2023a}]{hua2023up5}
Wenyue Hua, Yingqiang Ge, Shuyuan Xu, Jianchao Ji, and Yongfeng Zhang.
\newblock Up5: Unbiased foundation model for fairness-aware recommendation.
\newblock {\em arXiv preprint arXiv:2305.12090}, 2023.

\bibitem[\protect\citeauthoryear{Hua \bgroup \em et al.\egroup }{2023b}]{hua2023index}
Wenyue Hua, Shuyuan Xu, Yingqiang Ge, and Yongfeng Zhang.
\newblock How to index item ids for recommendation foundation models.
\newblock {\em arXiv preprint arXiv:2305.06569}, 2023.

\bibitem[\protect\citeauthoryear{J{\"a}rvelin and Kek{\"a}l{\"a}inen}{2002}]{NDCG}
Kalervo J{\"a}rvelin and Jaana Kek{\"a}l{\"a}inen.
\newblock Cumulated gain-based evaluation of ir techniques.
\newblock {\em ACM Transactions on Information Systems (TOIS)}, 20(4):422--446, 2002.

\bibitem[\protect\citeauthoryear{Jawahar \bgroup \em et al.\egroup }{2019}]{jawahar-etal-2019-bert}
Ganesh Jawahar, Beno{\^\i}t Sagot, and Djam{\'e} Seddah.
\newblock What does {BERT} learn about the structure of language?
\newblock In {\em Proceedings of the 57th Annual Meeting of the Association for Computational Linguistics}, pages 3651--3657, Florence, Italy, July 2019. Association for Computational Linguistics.

\bibitem[\protect\citeauthoryear{Kang and McAuley}{2018}]{kang2018self}
Wang-Cheng Kang and Julian McAuley.
\newblock Self-attentive sequential recommendation.
\newblock In {\em 2018 IEEE international conference on data mining (ICDM)}, pages 197--206. IEEE, 2018.

\bibitem[\protect\citeauthoryear{Li and Liang}{2021}]{li2021prefix}
Xiang~Lisa Li and Percy Liang.
\newblock Prefix-tuning: Optimizing continuous prompts for generation.
\newblock {\em arXiv preprint arXiv:2101.00190}, 2021.

\bibitem[\protect\citeauthoryear{Li \bgroup \em et al.\egroup }{2023a}]{recformer}
Jiacheng Li, Ming Wang, Jin Li, Jinmiao Fu, Xin Shen, Jingbo Shang, and Julian~J. McAuley.
\newblock Text is all you need: Learning language representations for sequential recommendation.
\newblock In {\em Proceedings of the 29th {ACM} {SIGKDD} Conference on Knowledge Discovery and Data Mining, {KDD} 2023, Long Beach, CA, USA, August 6-10, 2023}, pages 1258--1267, 2023.

\bibitem[\protect\citeauthoryear{Li \bgroup \em et al.\egroup }{2023b}]{li2023gpt4rec}
Jinming Li, Wentao Zhang, Tian Wang, Guanglei Xiong, Alan Lu, and Gerard Medioni.
\newblock Gpt4rec: A generative framework for personalized recommendation and user interests interpretation.
\newblock {\em arXiv preprint arXiv:2304.03879}, 2023.

\bibitem[\protect\citeauthoryear{Lu \bgroup \em et al.\egroup }{2020}]{lu2020twinbert}
Wenhao Lu, Jian Jiao, and Ruofei Zhang.
\newblock Twinbert: Distilling knowledge to twin-structured compressed bert models for large-scale retrieval.
\newblock In {\em Proceedings of the 29th ACM International Conference on Information \& Knowledge Management}, pages 2645--2652, 2020.

\bibitem[\protect\citeauthoryear{Ni \bgroup \em et al.\egroup }{2019}]{ni-etal-2019-justifying}
Jianmo Ni, Jiacheng Li, and Julian McAuley.
\newblock Justifying recommendations using distantly-labeled reviews and fine-grained aspects.
\newblock In {\em Proceedings of the 2019 Conference on Empirical Methods in Natural Language Processing and the 9th International Joint Conference on Natural Language Processing (EMNLP-IJCNLP)}, pages 188--197, Hong Kong, China, November 2019. Association for Computational Linguistics.

\bibitem[\protect\citeauthoryear{Rendle \bgroup \em et al.\egroup }{2010}]{rendle2010factorizing}
Steffen Rendle, Christoph Freudenthaler, and Lars Schmidt-Thieme.
\newblock Factorizing personalized markov chains for next-basket recommendation.
\newblock In {\em Proceedings of the 19th international conference on World wide web}, pages 811--820, 2010.

\bibitem[\protect\citeauthoryear{Rendle \bgroup \em et al.\egroup }{2012}]{rendle2012bpr}
Steffen Rendle, Christoph Freudenthaler, Zeno Gantner, and Lars Schmidt-Thieme.
\newblock Bpr: Bayesian personalized ranking from implicit feedback.
\newblock {\em arXiv preprint arXiv:1205.2618}, 2012.

\bibitem[\protect\citeauthoryear{Sanner \bgroup \em et al.\egroup }{2023}]{sanner2023cold}
Scott Sanner, Krisztian Balog, Filip Radlinski, Ben Wedin, and Lucas Dixon.
\newblock Large language models are competitive near cold-start recommenders for language- and item-based preferences, 2023.

\bibitem[\protect\citeauthoryear{Sun \bgroup \em et al.\egroup }{2019}]{sun2019bert4rec}
Fei Sun, Jun Liu, Jian Wu, Changhua Pei, Xiao Lin, Wenwu Ou, and Peng Jiang.
\newblock Bert4rec: Sequential recommendation with bidirectional encoder representations from transformer.
\newblock In {\em Proceedings of the 28th ACM international conference on information and knowledge management}, pages 1441--1450, 2019.

\bibitem[\protect\citeauthoryear{Tang and Wang}{2018}]{caser}
Jiaxi Tang and Ke~Wang.
\newblock Personalized top-n sequential recommendation via convolutional sequence embedding.
\newblock In {\em Proceedings of the eleventh ACM international conference on web search and data mining}, pages 565--573, 2018.

\bibitem[\protect\citeauthoryear{Tenney \bgroup \em et al.\egroup }{2019}]{tenney2019bert}
Ian Tenney, Dipanjan Das, and Ellie Pavlick.
\newblock Bert rediscovers the classical nlp pipeline.
\newblock {\em arXiv preprint arXiv:1905.05950}, 2019.

\bibitem[\protect\citeauthoryear{Wang \bgroup \em et al.\egroup }{2019}]{wang2019sequential}
Shoujin Wang, Liang Hu, Yan Wang, Longbing Cao, Quan~Z Sheng, and Mehmet Orgun.
\newblock Sequential recommender systems: challenges, progress and prospects.
\newblock In {\em arXiv preprint arXiv:2001.04830}, 2019.

\bibitem[\protect\citeauthoryear{Wang \bgroup \em et al.\egroup }{2023}]{wang2023generative}
Wenjie Wang, Xinyu Lin, Fuli Feng, Xiangnan He, and Tat-Seng Chua.
\newblock Generative recommendation: Towards next-generation recommender paradigm.
\newblock {\em arXiv preprint arXiv:2304.03516}, 2023.

\bibitem[\protect\citeauthoryear{Wu \bgroup \em et al.\egroup }{2017}]{wu2017recurrent}
Chao-Yuan Wu, Amr Ahmed, Alex Beutel, Alexander~J Smola, and How Jing.
\newblock Recurrent recommender networks.
\newblock In {\em wsdm}, 2017.

\bibitem[\protect\citeauthoryear{Xiao \bgroup \em et al.\egroup }{2023}]{bge_embedding}
Shitao Xiao, Zheng Liu, Peitian Zhang, and Niklas Muennighoff.
\newblock C-pack: Packaged resources to advance general chinese embedding, 2023.

\bibitem[\protect\citeauthoryear{Xie \bgroup \em et al.\egroup }{2022}]{xie2022contrastive}
Xu~Xie, Fei Sun, Zhaoyang Liu, Shiwen Wu, Jinyang Gao, Jiandong Zhang, Bolin Ding, and Bin Cui.
\newblock Contrastive learning for sequential recommendation.
\newblock In {\em 2022 IEEE 38th international conference on data engineering (ICDE)}, pages 1259--1273. IEEE, 2022.

\bibitem[\protect\citeauthoryear{Yu \bgroup \em et al.\egroup }{2024}]{yu2024ra}
Xiaohan Yu, Li~Zhang, Xin Zhao, Yue Wang, and Zhongrui Ma.
\newblock Ra-rec: An efficient id representation alignment framework for llm-based recommendation.
\newblock {\em arXiv preprint arXiv:2402.04527}, 2024.

\bibitem[\protect\citeauthoryear{Zhang \bgroup \em et al.\egroup }{2021}]{Zhang2021}
Yuhui Zhang, Hao Ding, Zeren Shui, Yifei Ma, James Zou, Anoop Deoras, and Hao Wang.
\newblock Language models as recommender systems: Evaluations and limitations.
\newblock In {\em NeurIPS 2021 Workshop on I (Still) Can't Believe It's Not Better}, 2021.

\bibitem[\protect\citeauthoryear{Zhang \bgroup \em et al.\egroup }{2022}]{zhang2022opt}
Susan Zhang, Stephen Roller, Naman Goyal, Mikel Artetxe, Moya Chen, Shuohui Chen, Christopher Dewan, Mona Diab, Xian Li, Xi~Victoria Lin, Todor Mihaylov, Myle Ott, Sam Shleifer, Kurt Shuster, Daniel Simig, Punit~Singh Koura, Anjali Sridhar, Tianlu Wang, and Luke Zettlemoyer.
\newblock Opt: Open pre-trained transformer language models, 2022.

\bibitem[\protect\citeauthoryear{Zhang \bgroup \em et al.\egroup }{2023}]{zhang2023instruction}
Shengyu Zhang, Linfeng Dong, Xiaoya Li, Sen Zhang, Xiaofei Sun, Shuhe Wang, Jiwei Li, Runyi Hu, Tianwei Zhang, Fei Wu, et~al.
\newblock Instruction tuning for large language models: A survey.
\newblock {\em arXiv preprint arXiv:2308.10792}, 2023.

\bibitem[\protect\citeauthoryear{Zhou \bgroup \em et al.\egroup }{2020}]{zhou2020s3}
Kun Zhou, Hui Wang, Wayne~Xin Zhao, Yutao Zhu, Sirui Wang, Fuzheng Zhang, Zhongyuan Wang, and Ji-Rong Wen.
\newblock S3-rec: Self-supervised learning for sequential recommendation with mutual information maximization.
\newblock In {\em Proceedings of the 29th ACM international conference on information \& knowledge management}, pages 1893--1902, 2020.

\end{thebibliography}

\end{document}